\documentclass[preprint,preprintnumbers,showpacs,aps,amssymb]{revtex4}

\usepackage{graphicx}
\usepackage{bm}
\usepackage{amsmath}

\begin{document}
\title{Comments on Verlinde's entropic gravity}
\author{Jong-Phil Lee}
\email{jplee@kias.re.kr}
\affiliation{Department of Physics and IPAP, Yonsei University, Seoul 120-749, Korea}
\affiliation{Division of Quantum Phases $\&$ Devices, School of Physics, Konkuk University, Seoul 143-701, Korea}

\begin{abstract}
Some comments are given on recently proposed entropic gravity by Verlinde.
We focus on the derivation of Newton's law of gravitation.
It is shown that consistent classical relations are enough to result in the Newtonian gravity.
In our derivation, it is crucial that the entropy is a quarter of the number of states in units of 
the Boltzmann constant $k_ B$, and the number of states is given by the holographic principle.
\end{abstract}

\maketitle
It has long been believed that gravity is one of the fundamental forces in nature since Newton.
But during the 1970s, there appeared some clues that gravity has something to do with the thermodynamics
\cite{Bekenstein,Hawking,Unruh}.
In 1995, Jacobson showed that the Einstein's field equations of general relativity are derived from the 
proportionality of entropy to the horizon area and the first law of thermodynamics \cite{Jacobson}.
Recently Padmanabhan used equipartition argument to provide a thermodynamic interpretation of gravity 
\cite{Padmanabhan}.
Soon after, Verlinde proposed that gravity is an entropic force caused by changes in the information
associated with the position of massive particles \cite{Verlinde}.
According to Verlinde, gravity is not a fundamental force but an emergent one.
After Verlinde, there have been a lot of works involving the idea of entropic force in cosmology \cite{cosmology}, 
black holes \cite{BH}, Rindler horizon \cite{Rindler}, corrections to the Newton's law \cite{corrections}, 
loop quantum gravity \cite{Smolin}, accelerating surfaces \cite{Makea},
etc.
There are also some critical views on Verlinde's approach \cite{critic}.
\par
The basic picture of Verlinde's proposition is that gravity arises from the entropic gradient $\Delta S$ over
the spatial region $\Delta x$ via the first law of thermodynamics
\begin{equation}
F=T\frac{\Delta S}{\Delta x}~,
\label{F}
\end{equation}
where $T$ is a temperature.
Verlinde argued that the change of entropy is associated with a particle of mass $m$ approaching to a
holographic screen with the displacement $\Delta x$.
Motivated by Bekenstein's entropy bound of black holes \cite{Bekenstein}, he postulated that the change of
entropy amounts to 
\begin{equation}
\Delta S=2\pi k_B \left(\frac{mc}{\hbar}\Delta x\right)~.
\label{DS}
\end{equation}
Note that $\lambda_C\equiv\hbar/mc$ is the Compton wavelength, and $\Delta x/\lambda_C$ counts the number
of states of the test particle when $\Delta x$ is comparable to $\lambda_C$.
(This was already pointed out in \cite{Padmanabhan}.)
The temperature comes from an acceleration $a$ through the Unruh's rule \cite{Unruh}
\begin{equation}
k_BT=\frac{1}{2\pi c}{\hbar} a~.
\label{Unruh}
\end{equation}
Plugging Eqs.\ (\ref{DS}) and (\ref{Unruh}) into Eq.\ (\ref{F}) yields the Newton's law, $F=ma$.
\par
On the other hand, if the energy is distributed over the holographic screen evenly,
the equipartition theorem tells that
\begin{equation}
E=\frac{1}{2}Nk_BT~,
\label{equipartition}
\end{equation}
where $N$ is the number of bits stored on the screen.
If a spherical holographic screen enclosing a mass $M$ is considered, 
one can easily get the Newton's law of gravitation, $F=GmM/r^2$, 
with the help of the holographic principle,
\begin{equation}
N=\frac{Ac^3}{G\hbar}~,
\label{N}
\end{equation}
where $A$ is the area of the holographic surface and $G$ is the Newton's constant, 
and with the relativistic energy $E=Mc^2$.
Verlinde extended his idea to the relativistic case to reach the Einstein's field equations of 
general relativity.
\par
But there are some aporiae in Verlinde's proposition.
First of all, the holographic screen feels relativistic energy of $Mc^2$ to provide a nonrelativistic
Newtonian gravitational force.
Secondly, the dependence of $\Delta S$ on $\Delta x$ in Eq.\ (\ref{DS}), 
not on the areal variation as is expected from the usual holographic principle,
does not seem to be reconciled with Eq.\ (\ref{N}).
Thirdly, it seems that Eq.\ (\ref{DS}) and the equipartition rule Eq.\ (\ref{equipartition}) is
inconsistent, because the thermodynamic relation $TS\sim E$ does not hold in this setup
\cite{critic}. 
Fourthly, the very concept of the holographic screen is rather obscure.
More explicitly, the increase of entropy must involve a lack of information.
In Bekenstein's case, this point is quite clear because the black hole horizon is a point of no return.
But in Verlinde's proposal, it is ambiguous which information is hidden when a test particle approaches
to the holographic screen.
If we believe both the second law of thermodynamics and Eq.\ (\ref{DS}), then the uncertainty of
$\Delta x$ becomes larger as the mass $m$ goes to $M$.
In other words, we cannot specify accurately the radial distance between the two masses as they get closer.
This is unnatural and counterintuitive.
\par
In this commentary note, we mainly concentrate on the derivation of Newton's law of gravitation.
Verlinde's starting point is the proportionality of entropy to the spatial displacement as given in Eq.\ (\ref{DS}).
But we propose that the conventional relation $S\sim N\sim A$ still reproduces the same result.
Let us first start with the usual entropy-area relation,
\begin{equation}
S=k_B\frac{1}{4}N=k_B\frac{Ac^3}{4G\hbar}~.
\label{S}
\end{equation}
When writing as above, one can immediately get $TS=\frac{1}{2}E$ \cite{Padmanabhan} for $E=\frac{1}{2}Nk_BT$ by equipartition rule.
For a spherical holographic screen of radius $r$ enclosing a mass $M$, the area is simply $A=4\pi r^2$, and so
\begin{equation}
\Delta S=k_B\frac{2\pi r c^3}{G\hbar}\Delta r~.
\end{equation}
The entropic force is given by
\begin{equation}
 F=T\frac{\Delta S}{\Delta r}=k_B T\frac{2\pi r c^3}{G\hbar}=\frac{2E}{N}\frac{2\pi r c^3}{G\hbar}~,
\end{equation}
where the equipartition rule is applied for $k_BT$.
At this point, Verlinde put $E=Mc^2$ to arrive at the Newton's law.
But $E$ is the energy stored on the holographic screen centered at $M$ with $m$ attached on its surface.
In nonrelativistic considerations, $E$ cannot be $Mc^2$.
A more reliable alternative to $Mc^2$ is just a classical potential energy $U(r)$ defined by $M$-$m$ system with the separation $r$.
The force $F$ is of course given by $-U'(r)$.
Thus we have a simple differential equation for $U(r)$:
\begin{equation}
\frac{dU}{U}+\frac{dr}{r}=0~.
\end{equation}
The solution is the familiar $U(r)=\frac{C}{r}$, reproducing Newton's inverse square law, where $C$ is a constant.
One drawback of this derivation is that the constant $C$ is not fixed, while in Verlinde's derivation the Newton's constant $G$
appears explicitly.
But our derivation is more consistent with all the known classical relations.
\par
As for derivation of the Einstein's equation of general relativity, there does not need to introduce 
the classical potential energy of $M$-$m$ system. 
One only needs the holographic screen enclosing a mass $M$ 
and an acceleration associated with the Unruh temperature which is in turn related to $M$ via equipartition rules.
Thus in the relativistic case it seems plausible to put $E=Mc^2$ as in \cite{Verlinde}.
\par
In the above manipulation, the fact that the entropy is a quarter of the number of states in units of $k_B$ is crucial to yield the
inverse square law.
In general, if the entropy is given by
\begin{equation}
S=\gamma k_B N~,
\end{equation}
then the potential $U(r)$ satisfies
\begin{equation}
\frac{1}{U}\frac{\Delta U}{\Delta r}+2\gamma\frac{1}{N}\frac{\Delta N}{\Delta r}=0~,
\end{equation}
and thus
\begin{equation} 
 U=\frac{C_N}{N^{2\gamma}}~,
\end{equation}
where $C_N$ is an integration constant.
If $N$ scales as $N\sim r^\alpha$, then 
\begin{equation}
U(r)=\frac{C_r}{r^{2\gamma\alpha}}~,
\end{equation}
for some constant $C_r$.
Therefore, the holographic principle ($\alpha=2$) and ''a quarter ($\gamma=1/4$)'' in the entropy inevitably result in the Newtonian gravity.
If information is stored in a 'voxelated' (not 'pixelated' as in the holography) form (i.e., $\alpha=3$), then we must have a non-Newtonian
potential for $\gamma=1/4$, or $\gamma=1/6$ to keep the inverse square law.
Or, if we reverse the logic, for a given Newtonial potential, $\gamma=1/4$ requires the holographic principle, and vice versa.
\par
In our framework, it is not likely to have a meaningful Unruh temperature because current examination is classical. 
To get the Unruh effect, one should introduce an accelerating system which needs relativistic analysis.
Also, it is still unclear which information is hidden when entropic gradient causes gravity.
For example, if the number of bits $N$ for $M$-$m$ system with separation $r$ is given by Eq.\ (\ref{N}) 
where $A=4\pi r^2$,
then $N$ decreases when $M$ and $m$ approach each other, apparently violating the second law of
thermodynamics.
So there must be some other kinds of entropy increase during the process, 
or the entropy of the system should be defined differently.
Unfortunately, there are no known alternatives yet.
\par
In conclusion, we have investigated the Verlinde's entropic force as the origin of gravity.
It was shown that there are some inconsistencies in Verlinde's arguements in a classical point of view.
Furthermore, the change of entropy proportional to the displacement of a test particle is controversial.
We found that if it follows the conventional areal proportionality and the energy stored on the holographic
surface is identified as the classical potential,
then the Newtonian law of gravitation comes out as an entropic force.
\par
But there are still some issues to be clarified.
Above all, one should make it clear how the number of states (or, the entropy) increases 
when there are two masses, or how they feel the entropic gradient to pull together toward each other.
\begin{acknowledgments}
This work was supported by the Basic Science Research Program through the National Research Foundation of Korea (NRF) 
funded by the Korean Ministry of Education, Science and Technology (2009-0088396).
\end{acknowledgments}


\begin{thebibliography}{99}
\bibitem{Bekenstein}
  J.~D.~Bekenstein,
  Phys.\ Rev.\  D {\bf 7}, 2333 (1973).
\bibitem{Hawking}
  S.~W.~Hawking,
  Commun.\ Math.\ Phys.\  {\bf 43}, 199 (1975)
  [Erratum-ibid.\  {\bf 46}, 206 (1976)].
\bibitem{Unruh}
  W.~G.~Unruh,
  Phys.\ Rev.\  D {\bf 14}, 870 (1976).
\bibitem{Jacobson}
  T.~Jacobson,
  Phys.\ Rev.\ Lett.\  {\bf 75}, 1260 (1995)
  [arXiv:gr-qc/9504004].
\bibitem{Padmanabhan}
  T.~Padmanabhan,
  arXiv:0912.3165 [gr-qc].
\bibitem{Verlinde}
  E.~P.~Verlinde,
  arXiv:1001.0785 [hep-th].
\bibitem{cosmology}
 F.~W.~Shu and Y.~Gong,
  arXiv:1001.3237 [gr-qc];
R.~G.~Cai, L.~M.~Cao and N.~Ohta,
  Phys.\ Rev.\  D {\bf 81}, 061501 (2010)
  [arXiv:1001.3470 [hep-th]];
 M.~Li and Y.~Wang,
  Phys.\ Lett.\  B {\bf 687}, 243 (2010)
  [arXiv:1001.4466 [hep-th]];
Y.~Zhang, Y.~g.~Gong and Z.~H.~Zhu,
  arXiv:1001.4677 [hep-th];
Y.~Wang,
  arXiv:1001.4786 [hep-th];
S.~W.~Wei, Y.~X.~Liu and Y.~Q.~Wang,
  arXiv:1001.5238 [hep-th];
 Y.~Ling and J.~P.~Wu,
  arXiv:1001.5324 [hep-th];
J.~W.~Lee, H.~C.~Kim and J.~Lee,
  arXiv:1001.5445 [hep-th];
D.~A.~Easson, P.~H.~Frampton and G.~F.~Smoot,
  arXiv:1002.4278 [hep-th],
  arXiv:1003.1528 [hep-th];
 U.~H.~Danielsson,
  arXiv:1003.0668 [hep-th];
J.~W.~Lee,
  arXiv:1003.1878 [hep-th];
Y.~F.~Cai, J.~Liu and H.~Li,
  arXiv:1003.4526 [astro-ph.CO];
 A.~Sheykhi,
  arXiv:1004.0627 [gr-qc];
M.~Li and Y.~Pang,
  arXiv:1004.0877 [hep-th].
\bibitem{BH}
 Y.~S.~Myung,
  arXiv:1002.0871 [hep-th];
Y.~X.~Liu, Y.~Q.~Wang and S.~W.~Wei,
  arXiv:1002.1062 [hep-th];
R.~G.~Cai, L.~M.~Cao and N.~Ohta,
  Phys.\ Rev.\  D {\bf 81}, 084012 (2010)
  [arXiv:1002.1136 [hep-th]].
Y.~Tian and X.~Wu,
  arXiv:1002.1275 [hep-th].
\bibitem{Rindler}
H.~Culetu,
  arXiv:1001.4740 [hep-th],
  arXiv:1002.3876 [hep-th];
 J.~W.~Lee,
  arXiv:1003.4464 [hep-th].
\bibitem{corrections}
C.~Gao,
  arXiv:1001.4585 [hep-th];
 S.~Ghosh,
  arXiv:1003.0285 [hep-th];
 L.~Modesto and A.~Randono,
  arXiv:1003.1998 [hep-th].
M.~R.~Setare and D.~Momeni,
  arXiv:1004.2794 [physics.gen-ph].
\bibitem{Smolin}
L.~Smolin,
  arXiv:1001.3668 [gr-qc].
\bibitem{Makea}
 J.~Makea,
  arXiv:1001.3808 [gr-qc].
\bibitem{critic}
 S.~Gao,
  arXiv:1002.2668 [gr-qc];
see also the second paper of \cite{Rindler}.
\end{thebibliography}
\end{document}